\documentclass[iop,apj,tighten]{emulateapj}
\usepackage{amsmath,amstext}
\usepackage{apjfonts}
\usepackage[breaklinks,colorlinks,citecolor=blue,linkcolor=magenta]{hyperref} 
\usepackage[all]{hypcap} %Links go to figures. This breaks deluxetables; use \capstartfalse \capstarttrue around deluxetables to fix it.
\usepackage{float}
\usepackage{footmisc}
\usepackage{multirow}
\usepackage{lineno}
\usepackage{natbib}

\def \nustar {{\em NuSTAR}}

\def \chandra {{\em Chandra}}
\def \nicer {{\em NICER}}

\def \swift {{\em Swift}}

\def \maxi {{MAXI~J1535-571}}

\shorttitle{\maxi}
\shortauthors{Xu et al.}

\begin{document}

\title{Reflection Spectra of the Black Hole Binary Candidate \maxi\ in the Hard State observed by NuSTAR}

\author{Yanjun Xu\altaffilmark{1}}
\author{Fiona A. Harrison\altaffilmark{1}}
\author{Javier A. Garc\'ia\altaffilmark{1,2}}
\author{Andrew C. Fabian\altaffilmark{3}}
\author{Felix F\"urst\altaffilmark{4}}
\author{Poshak Gandhi\altaffilmark{5}}
\author{Brian W. Grefenstette\altaffilmark{1}}
\author{Kristin K. Madsen\altaffilmark{1}}
\author{Jon M. Miller\altaffilmark{6}}
\author{Michael L. Parker\altaffilmark{4}}
\author{John A. Tomsick\altaffilmark{7}}
\author{Dominic J. Walton\altaffilmark{3}}

\altaffiltext{1}{Cahill Center for Astronomy and Astrophysics, California Institute of Technology, Pasadena, CA 91125, USA}
\altaffiltext{2}{Remeis Observatory \& ECAP, Universit\"at Erlangen-N\"urnberg, Sternwartstr.~7, 96049 Bamberg, Germany}
\altaffiltext{3}{Institute of Astronomy, University of Cambridge, Madingley Road, Cambridge CB3 0HA, UK}
\altaffiltext{4}{European Space Astronomy Centre (ESA/ESAC), Operations Department, Villanueva de la Ca\~nada (Madrid), Spain}
\altaffiltext{5}{Department of Physics and Astronomy, University of Southampton, SO17 3RT, UK}
\altaffiltext{6}{Department of Astronomy, University of Michigan, 1085 South University Avenue, Ann Arbor, MI 48109, USA}
\altaffiltext{7}{Space Sciences Laboratory, 7 Gauss Way, University of California, Berkeley, CA 94720-7450, USA}
\altaffiltext{7}{European Space Astronomy Centre (ESA/ESAC), Operations Department, Villanueva de la Ca\~nada (Madrid), Spain}

\begin{abstract}
We report on a \nustar\ observation of the recently discovered bright black hole candidate \maxi. \nustar\ observed the source on MJD 58003 (five days after the outburst was reported). The spectrum is characteristic of a black hole binary in the hard state. We observe clear disk reflection features, including a broad Fe K$\alpha$ line and a Compton hump peaking around 30~keV. Detailed spectral modeling reveals  narrow Fe K$\alpha$ line complex centered around 6.5~keV on top of the strong relativistically broadened Fe K$\alpha$ line. The narrow component is consistent with distant reflection from moderately ionized material. The spectral continuum is well described by a combination of cool thermal disk photons and a Comptonized plasma with the electron temperature $kT_{\rm e}=19.7\pm{0.4}$~keV. An adequate fit can be achieved for the disk reflection features with a self-consistent relativistic reflection model that assumes a lamp-post geometry for the coronal illuminating source. The spectral fitting measures a black hole spin $a>0.84$, inner disk radius $R_{\rm in}<2.01~r_{\rm ISCO}$, and a lamp-post height $h=7.2^{+0.8}_{-2.0}~r_{\rm g}$ (statistical errors, 90\% confidence), indicating no significant disk truncation and a compact corona. Although the distance and mass of this source are not currently known, this suggests the source was likely in the brighter phases of the hard state during this \nustar\ observation.

\end{abstract}
 
\keywords{accretion, accretion disks --- black hole physics --- X-rays: binaries} 
\maketitle

%%%%%%%%%%%%%%%%%%%%%%%%%%
\section{INTRODUCTION}
%%%%%%%%%%%%%%%%%%%%%%%%%%
\maxi\ was discovered by MAXI/GSC \citep{nego17} and Swift/BAT \citep{kenn17a} as an uncatalogued hard X-ray transient located near the Galactic plane on September 2, 2017. Subsequent monitoring in X-ray and radio indicates behavior consistent with other known black hole transients, making it a strong black hole binary candidate \citep[e.g.,][]{nego17, maxi_atca}. The optical and near-infrared counterparts were identified \citep{dincer17,scar17}. The source was reported to begin the hard-to-soft state transition around September 10 \citep{kenn17b,nak17}, soon followed by the detection of low-frequency quasi-periodic oscillations (QPOs) by \swift/XRT \citep{mere17}. The X-ray flux level of \maxi\ rose rapidly, reaching $\sim$5 Crab in the MAXI/GSC band, making it one of the brightest black hole binary candidates known. At the time of this work the source is still in outburst and continues to evolve.

During a typical outburst, black hole binaries undergo a transition from the low/hard to the high/soft state through relatively short-lived intermediate states \citep[see][for a review]{bhb_rev06}. This process is believed to be associated with changes in the accretion flow geometry at the vicinity of the black hole. It is generally well accepted that the inner disk extends to the innermost stable circular orbit (ISCO) in the soft state. Since the black hole angular momentum sets the location of the ISCO radius, estimations of the black hole spin are possible via X-ray spectroscopy, which has been achieved either by modeling the thermal disk \citep[e.g.,][]{zhang97} or the disk reflection component \citep[e.g.,][]{fab89}. In addition, from the relativistic reflection spectrum, information can be obtained about the nature of the illuminating source commonly referred to as the ``corona" \citep[e.g.,][]{miller15,dom17v404}.

The interpretation of the hard state spectrum is still highly debated. The disk accretion model given by \cite{esin97} suggests the inner disk is truncated and replaced by an advection-dominated accretion flow at low mass accretion rates in the low/hard state. A recessed disk has also been invoked to explain the behavior of low-frequency QPOs commonly found in the hard state \citep[e.g.,][]{ingram09,ingram16}. There have been efforts to measure the inner disk radius in the hard state through reflection modeling \citep[e.g.,][]{furst15,gar15}. However, controversies remain about the radius of truncation and when disk truncation occurs in terms of the Eddington ratio. The results in some cases have been questioned because of photon pile-up issues at high count rates \citep[e.g.,][]{done10,miller10}.

With high sensitivity, broad bandpass and triggered read-out free from pile-up distortion, the \textit{Nuclear Spectroscopic Telescope Array} \citep[\nustar,][]{nustar} is ideal for studying reflection in Galactic binaries. Recent \nustar\ observations of several black hole binaries in the bright hard state revealed very broad iron lines \citep[][]{miller13,miller15}, which are among the best evidence for lack of disk truncation. New observations of black hole binaries in outburst are important for understanding the accretion geometry in the hard state.

%%%%%%%%%%%%%%%%%%%%%%%%%%
\section{OBSERVATION AND DATA REDUCTION}
%%%%%%%%%%%%%%%%%%%%%%%%%%
\label{sec:data}
\maxi\ was observed by \nustar\ \citep{nustar} starting on 2017 September 7 (MJD 58003) at 18:41:09 UT under a DDT request (OBSID 90301013002). We processed the \nustar\ data using v.1.6.0 of the NuSTARDAS pipeline with \nustar\ CALDB v20170817. After filtering background flares due to enhanced solar activity by setting {\tt saacalc=2}, {\tt saamode} = {\tt OPTIMIZED} and {\tt tentacle = no} in NUPIPELINE, and the correction for dead time, the effective exposure times are 8.7~ks and 9.1~ks for the two focal plane modules FPMA and FPMB, respectively. The spectra were extracted from a circular region of the radius 180$\arcsec$ centered on the source location. We chose the background from a blank region on the detector furthest from source location to avoid source photons. The spectra were grouped to have a signal-to-noise (S/N) ratio of at least 30 per bin.

%%%%%%%%%%%%%%%%%%%%%
%%%%%%%------figure1--------%%%%%%%
\begin{figure}
\centering
\includegraphics[width=0.49\textwidth]{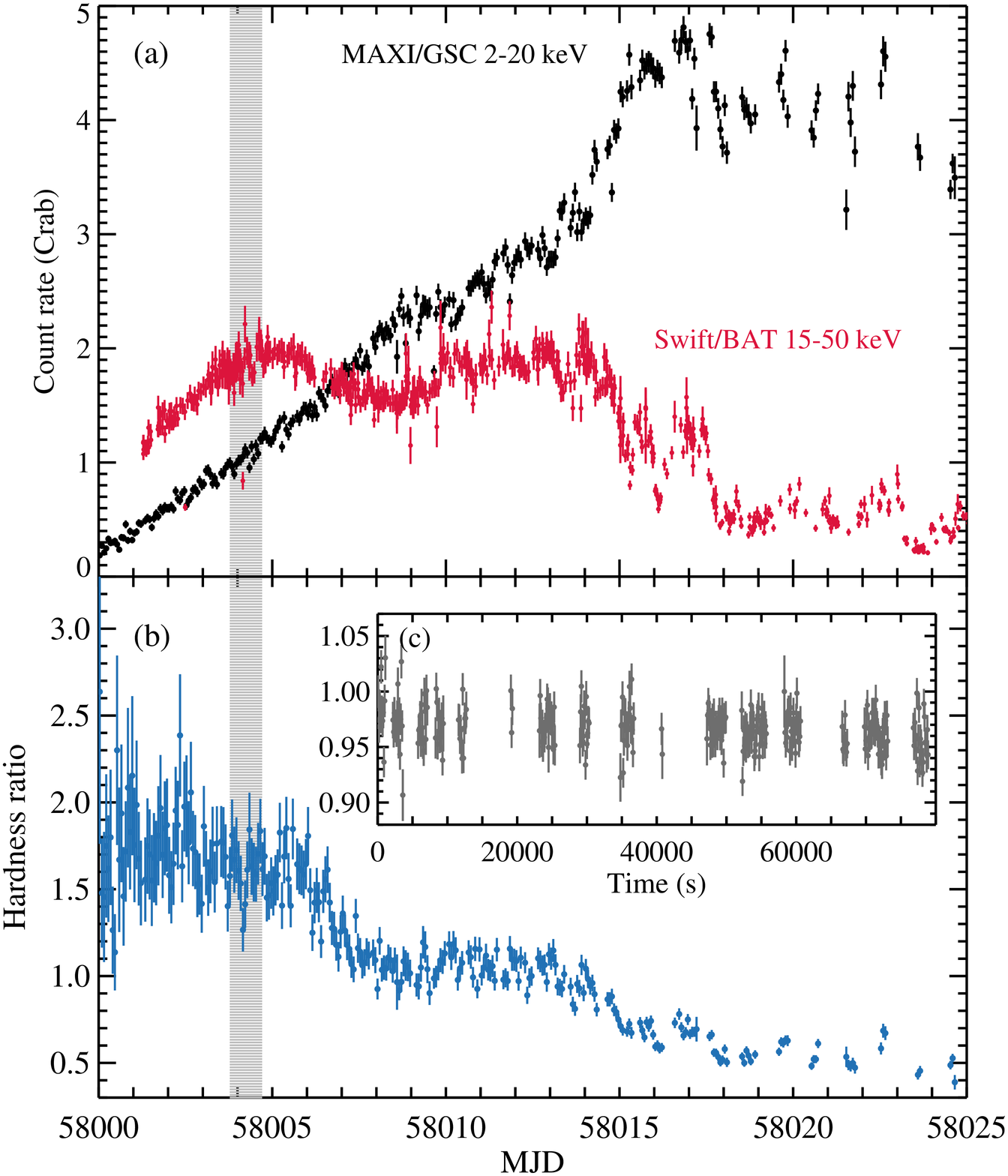}
\caption{(a) MAXI/GSC (black) and Swift/BAT (red) orbital light curves of \maxi, scaled to the Crab count rates in the corresponding instrument bands (only BAT data points with S/N $>$ 7 are included). The gray shaded area marks the duration of the \nustar\ observation. (b) MAXI hardness ratio calculated from count rates (4--20 / 2--4~keV).  (c) \nustar\ hardness ratio in 100s bins (count rates, 6--10 / 3--6 keV).
\label{fig:fig1}}
\end{figure}
 
As shown in Figure~\ref{fig:fig1}, the \nustar\ observation caught \maxi\ in the hard state before the spectrum began to soften. The source flux rose quickly during the observation, the dead time corrected FPMA count rate increased from $\sim$750~cts~s$^{-1}$ to $\sim$900~cts~s$^{-1}$ from the start to the end of the exposure, exceeding the Crab count rate ($\sim$500~cts~s$^{-1}$) in the \nustar\ band \citep{mad_crab15}. We only consider time-averaged spectra in this work, as there is no significant change in the hardness ratio (Figure~\ref{fig:fig1}c). We model the \nustar\ spectra using XSPEC v12.9.0n \citep{xspec} using $\chi^{2}$ statistics, and adopt the cross-sections from \cite{crosssec} and abundances from \cite{wil00}. All parameter uncertainties are reported at the 90\% confidence level for one parameter of interest. A cross-normalization constant is allowed to vary freely for FPMB and is assumed to be unity for FPMA.

%%%%%%%%%%%%%%%%%%%%%%%%%%
\section{SPECTRAL MODELING}
%%%%%%%%%%%%%%%%%%%%%%%%%%
\label{sec:ana}

\begin{figure}
\centering
\includegraphics[width=0.49\textwidth]{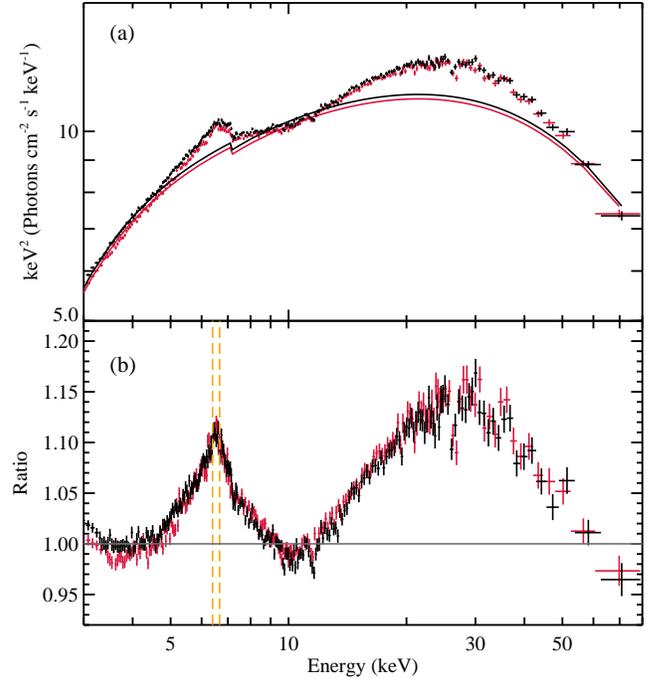}
\caption{(a) Unfolded \nustar\ spectra of \maxi. FPMA and FPMB data are plotted in black and red, respectively. (b) Data/model residuals to an absorbed cutoff power-law model. The narrow core of the iron line peaks at $\sim$6.5~keV. The energies of neutral Fe K${\alpha}$ (6.4~keV) and Fe~{\scriptsize XXV} K${\alpha}$ (6.7~keV) are marked in orange dashed lines. \label{fig:fig2}}
\end{figure}

A clear reflection component is present in the \nustar\ spectra. To highlight the reflection features, we first fit the spectra with an absorbed cutoff power-law model, {\tt TBabs*cutoffpl} in XSPEC notation, only considering the energy intervals of 3--4, 8--12, 40--79~keV. This approximate fit requires a power-law index $\Gamma\sim$1.6 and high-energy cutoff at $E_{\rm cut}\sim$60~keV. As displayed in Figure~\ref{fig:fig2}, a broad iron line extending down to about 4.5~keV and a Compton hump peaking around 30 keV is evident. In addition, a sharp Fe K-edge can be seen at 7.1~keV, indicating strong absorption.

%%%%%%%------figure2--------%%%%%%%

In the residuals, we also notice narrow dips in the spectra at 11.5 and 26.5~keV, which would only be obvious in bright sources with high S/N data. Whether these are instrumental related is unclear and currently under investigation. We ignore the corresponding energy bins in the spectral fitting from here on. There is also a slight difference ($\sim$1\%) between FPMA and FPMB below 5~keV as can be seen in Figure~\ref{fig:fig2}b, which is within the calibration accuracy of \nustar\ \citep{madsen15}. In order to improve the fitting statistics, we account for this discrepancy by fitting the photon-index, $\Gamma$, of FPMA and FPMB independently. In all fits in this work, the difference in $\Gamma$ is minimal (within 0.01) and would not influence other parameters. Therefore we only report the value of $\Gamma$ from FPMA. 

Motivated by the low high-energy cutoff, we model the continuum with the thermal Comptonization model {\tt nthcomp} \citep{zdz96,zyc99}. The shape of thermal Comptonization is significantly different from a cutoff power-law even below 80 keV \citep{zdz03}. To minimize the number of free parameters, we first fit the disk reflection assuming a ``lamp-post" geometry, where the corona is a point source located on the spin axis of the black hole at a height, $h$, above the accretion disk. The source emissivity profile can be self-consistently calculated with the lamp-post assumption given the location of the illuminating source \citep{dauser13}. We use the lamp-post model {\tt relxilllpCp} in the {\tt relxill} model family \citep{relxilla, relxillb}, which internally includes a {\tt nthcomp} continuum. The reflection fraction can be self-consistently determined in {\tt relxilllpCp} based on the inner disk radius $R_{\rm in}$, the spin parameter $a$ and the lamp-post height $h$ from ray-tracing calculations, which helps to constrain the geometry of the system and rules out some unphysical parts of the parameter space \citep[see][for a discussion]{relxilla}. We fit for $a$ and $R_{\rm in}$ simultaneously, and freeze the outer edge of accretion disk, $R_{\rm out}$, at 400~$r_{\rm g}$, where $r_{\rm g}$ is the gravitational radius defined as $r_{\rm g} \equiv {{\rm GM}/c^2}$). In addition, we include possible contribution from reprocessing by distant material using an unblurred reflection model {\tt xillverCp} \citep{garcia10} to account for the narrow core of the Fe~K$\alpha$ line. We first assume the distant reprocessing to be neutral by fixing log~${\xi}=0$ (where the ionization parameter $\xi \equiv L/nR^2$, $L$ is the ionizing luminosity, $n$ is the gas density and $R$ is the distance to the ionizing source), as neutral narrow Fe~K$\alpha$ lines have been commonly found in bright Galactic binaries \citep[e.g.,][]{parker15,walton16}. The iron abundance $A_{\rm Fe}$ and the input continuum in {\tt xillverCp} are linked with those in the disk reflection component. The total model setup is {\tt TBabs*(relxilllpCp+xillverCp)} (Model 1).

\begin{figure*}
\centering
\includegraphics[width=0.98\textwidth]{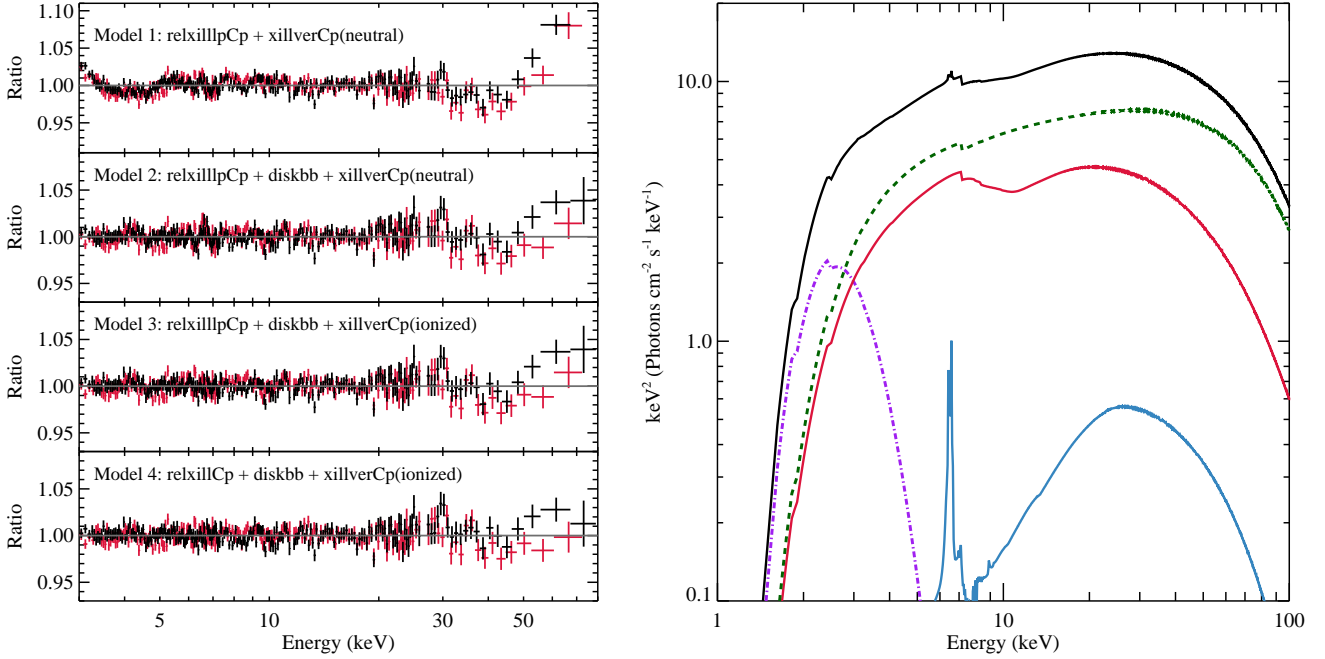}
\caption{Left: Ratio plots of the reflection modeling. The spectra are rebinned for display clarity. Right: Contributions from different spectral components in Model 3. The total model is marked in black solid lines, together with the thermal disk (purple), Comptonization continuum (dark green), disk reflection (red) and distant reflection (blue). 
\label{fig:fig3}}
\end{figure*}

As shown in Figure~\ref{fig:fig3} (left panel), Model 1 fails to adequately fit the spectral continuum, leaving obvious excesses at both the soft and hard end of the \nustar\ energy band. The reduced $\chi^2_{\nu}$ ($\chi^2/\nu$ where $\nu$ is the number of degrees of freedom) is $1986/1371=1.45$. The fit can be greatly improved by adding a multi-color disk blackbody component: {\tt TBabs*(relxilllpCp+xillverCp+diskbb)} (Model 2), with $\chi^2/\nu=1586/1369=1.16$. A disk component is not evident in the \swift/XRT data as of September 11 \citep{kenn17b}, which may be due to the high level of obscuration. However, we note that the contribution from the thermal disk is important for an adequate fit even at higher energies, because it allows for a harder continuum. There are still some residuals left, a narrow peak between 6 to 7~keV (see Figure~\ref{fig:fig3}). As can be seen in Figure~\ref{fig:fig2}b, the narrow core of the iron line actually peaks between 6.4~keV (neutral Fe K$\alpha$) and 6.7 keV (Fe {\small XXV} K${\alpha}$), indicating that the distant reprocessing is most likely ionized. Therefore, we leave the ionization parameter to vary freely in {\tt xillverCp} (Model 3).

%%%%%%table%%%%%%%%%%%%%%
%%%%%%%------table--------%%%%%%%
\capstartfalse
\begin{deluxetable}{clcc}
\tablewidth{\columnwidth}
\tablecolumns{6}
\tabletypesize{\scriptsize}
\tablecaption{Best-fit Model Parameters \label{tab:tab1}}
\tablehead{
\colhead{Component} & \colhead{Parameter} &\colhead{Model 3} &\colhead{Model 4}} 
\startdata

{\textsc{tbabs}}     & $N_{\rm H}$ ($\rm \times10^{22}~cm^{-2}$) &$8.2^{+0.3}_{-0.6}$ & $7.2\pm{0.3}$  \\
\noalign{\smallskip}
{\textsc{diskbb}}    & $kT_{\rm in}$ (keV)  &$0.43\pm{0.01}$   &$0.40\pm{0.01}$ \\
\noalign{\smallskip}
                     & {Norm ($10^5$)}   & $1.07^{+0.20}_{-0.07}$   & $1.2^{+0.6}_{-0.3}$     \\
\noalign{\smallskip}
{\textsc{relxill(lp)Cp}} & $h$ ($r_{\rm g}$) & $7.2^{+0.8}_{-2.0}$ & \nodata  \\
\noalign{\smallskip} 
				     & $q_{\rm in}$  &\nodata & $>9.2$  \\
\noalign{\smallskip} 
				     & $q_{\rm out}$  &\nodata & $3^{*}$  \\
\noalign{\smallskip} 
				     & $R_{\rm br}$ ($r_{\rm g}$)  & \nodata & $10^{*}$  \\
\noalign{\smallskip} 
				     & $a$ ($c{\rm J/GM}^{2}$) & $>0.84$  &$>0.987$ \\
\noalign{\smallskip}  
                     & $R_{\rm in}$ ($r_{\rm ISCO}$) & $<2.01$  & $<1.22$\\
\noalign{\smallskip}
				     & $i$ ($^\circ$)   & $57^{+1}_{-2}$  &$75^{+2}_{-4}$  \\
\noalign{\smallskip}                         
				     & $\Gamma$   & $1.815^{+0.005}_{-0.008}$ &$1.862^{+0.014}_{-0.016}$ \\
\noalign{\smallskip}                        
                      & log~${\xi}$ (log [$\rm erg~cm~s^{-1}$]) & $3.69\pm{0.04}$  & $3.19^{+0.21}_{-0.15}$\\
\noalign{\smallskip}                          
				      & {$A_{\rm Fe}$ (solar)}  &$1.4^{+0.3}_{-0.1}$  &$0.8\pm{0.1}$ \\
\noalign{\smallskip}                          
				      &  {$kT_{\rm e}$~(keV)}  & $19.7\pm{ 0.4}$  &$21.9\pm{1.2}$ \\
\noalign{\smallskip}  
                      &  $R_{\rm ref}$  & 1.55 & $0.60^{+0.06}_{-0.10}$   \\
\noalign{\smallskip}  
                     & {Norm}   & $0.129^{+0.006}_{-0.009}$  & $0.089^{+0.007}_{-0.008}$      \\
\noalign{\smallskip}
{\textsc{xillverCp}}  & log~${\xi}$ (log [$\rm erg~cm~s^{-1}$])  & $2.35^{+0.10}_{-0.08}$ & $2.32^{+0.10}_{-0.14}$  \\
\noalign{\smallskip}
    & {Norm ($10^{-3}$)}   & $7.8^{+1.0}_{-0.7}$  & $17^{+3}_{-5}$     \\
\noalign{\smallskip}
\hline	
\noalign{\smallskip}		
            &   $\chi^2/{\nu}$  & $1538/1368$  & $1515/1367$\\
\noalign{\smallskip}
\hline
\noalign{\smallskip}
\multicolumn{2}{c}{F$_{\rm 3-10~keV}$~(erg~cm$^{-2}$~s$^{-1}$)~$^{\rm a}$}   & \multicolumn{2}{c}{$1.67\times10^{-8}$}\\
\multicolumn{2}{c}{F$_{\rm 10-79~keV}$~(erg~cm$^{-2}$~s$^{-1}$)~$^{\rm b}$} & \multicolumn{2}{c}{$3.62\times10^{-8}$} \\
\multicolumn{2}{c}{F$_{\rm 0.1-500~keV}$~(erg~cm$^{-2}$~s$^{-1}$)~$^{\rm c}$} & $8.10\times10^{-8}$ & $7.82\times10^{-8}$
\enddata
\tablecomments{
Frozen parameters are marked with asterisks. There is no error estimation for $R_{\rm ref}$ in Model 3, as the parameter is self-consistently calculated in the lamp-post geometry. $^{\rm a,b}$ Absorbed flux calculated from the normalization of FPMA. $^{\rm c}$ Unabsorbed flux.
}
\end{deluxetable}

%%%%%%%%%%%%%%%%%%%%%
%%%%%%%------figure3--------%%%%%%%

%%%%%%%%%%%%%%%%%%%%%
%%%%%%%------figure4--------%%%%%%%
\begin{figure*}
\centering
\includegraphics[width=0.75\textwidth]{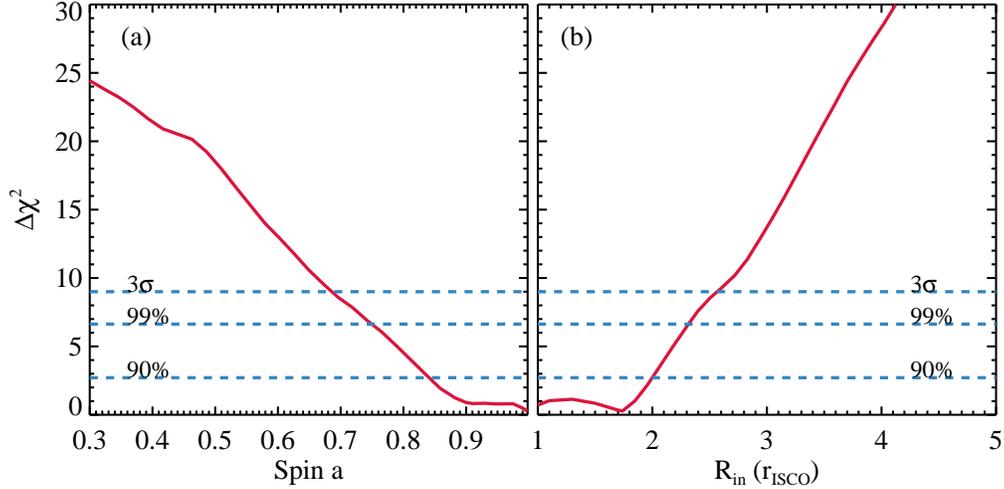}
\caption{$\Delta\chi^2$ plots for the black hole spin parameter a and inner accretion disk radius $R_{\rm in}$ from Model 3. The dashed lines mark the 90\%, 99\% and $3\sigma$ confidence levels for one parameter of interest.
\label{fig:fig4}}
\end{figure*}

Model 3 yields a reasonable fit, $\chi^2/\nu=1538/1368=1.12$ with no obvious residuals (see Figure~\ref{fig:fig3}). Allowing the distant reflection component to be ionized brings an improvement of $\Delta\chi^2=48$ for one additional parameter, with the ionization parameter for the distant reprocessing material measured as log~${\xi}=2.35^{+0.10}_{-0.08}$, indicating that the narrow core is actually a blend of Fe K line complex (which are not well separated by \nustar). The relative contributions from different components are plotted in Figure~\ref{fig:fig3} (right panel). The Comptonization model {\tt nthcomp} describes the incident continuum by the asymptotic photon-index, $\Gamma$, and the electron temperature, $kT_{e}$, which are well measured to be $1.815^{+0.005}_{-0.008}$ and $19.7\pm0.4$~keV, respectively. We note the value of $kT_{e}$ here is as observed, not in the source frame. The model also finds an inclination angle of $i=57^{+1}_{-2}$$^{\circ}$, an iron abundance $A_{\rm Fe}=1.4^{+0.3}_{-0.1}$ (in solar units), and an accretion disk ionization parameter log~${\xi}=3.69\pm{0.04}$ (see Table~\ref{tab:tab1}).

In addition, the disk reflection modeling indicates a relatively low lamp-post height of $h=7.2^{+0.8}_{-2.0}~r_{\rm g}$. The constraints on the spin, $a$, and inner disk radius, $R_{\rm in}$, are plotted in Figure~\ref{fig:fig4}. Here we measure $R_{\rm in}$ in units of $r_{\rm ISCO}$ to better test whether the disk is truncated as the black hole spin is not known a priori. The radius of the ISCO is a function of the black hole spin, decreasing monotonically from 6 $r_{\rm g}$ for a Schwarzschild black hole to 1.235 $r_{\rm g}$ for a black hole with the maximum spin parameter of 0.998. The results are strongly inconsistent with a significantly truncated accretion disk scenario, with the best-fit parameters $a>0.84$ and $R_{\rm in}<2.01~r_{\rm ISCO}$. The $\Delta\chi^2$ contour flattens down at the regime of high $a$ and low $R_{\rm in}$ (Figure~\ref{fig:fig4}), which is expected as the two parameters are degenerate, both controlling the absolute position of the inner disk radius. The reflection fraction $R_{\rm ref}$ given by the lamp-post model is 1.55, defined to be the ratio of the coronal intensity illuminating the disk to that reaching the observer \citep{dauser16}. A high reflection fraction is an indicator of strong light bending effects, resulting from a combination of a low corona height, a small inner disk radius and a high spin. Fitting for the reflection fraction as a free parameter does not result in a significantly improved fit.

The model also confirms the obscured nature of the source with the absorption column density $N_{\rm H}=8.2^{+0.3}_{-0.6}\times 10^{22}$~cm$^{-2}$. The value is higher than that reported by \swift/XRT \citep[$N_{\rm H}=3.6\pm0.2\times 10^{22}$~cm$^{-2}$,][]{kenn17b}, but a higher $N_{\rm H}$ is expected in order to obtain a similar spectral shape with an extra thermal disk component included; a high $N_{\rm H}$ was also reported by preliminary results from \nicer\ \citep[$N_{\rm H}=4.89\pm0.06\times 10^{22}$~cm$^{-2}$,][]{nicer17}. We note that the values of $N_{\rm H}$ reported here were measured at different epochs, thus the difference could possibly be due to variations in the intrinsic absorption column during the outburst \citep[e.g.,][]{dom17v404}. The inner disk temperature measured is $kT_{\rm in}=0.43\pm{0.01}~{\rm keV}$, which is somewhat higher than the values of several well-known black hole binaries in the hard state \citep[e.g.,][]{rey10, basak16}. An alternative interpretation to the soft emission is to invoke a soft Comptonization component \citep[e.g.,][]{salvo01}. We explored this possibility by including an extra {\tt nthcomp} model following \cite{basak17}, but it brings no further improvement to the fit and causes no significant change in other parameters. In addition, we notice that the low-energy end of the spectra includes minor contributions from a dust scattering halo. Therefore, we stress that without soft band coverage, an accurate and unbiased determination of the disk temperature and absorption column density is difficult, but this uncertainty has negligible influence on the reflection parameters as the spectra can be well fitted above 5 keV without a thermal disk or soft Comptonization component, and the other parameters remain unchanged.

To explore the disk emissivity beyond the assumption from a lamp-post geometry, we use a broken power-law emissivity profile ($\epsilon(r)\varpropto r^{-q}$) in the {\tt relxillCp} model, which is described by the inner and outer emissivity indexes $q_{\rm in, out}$ and a break radius $R_{\rm br}$ (Model 4). The best-fit inner emissivity index, $q_{\rm in}$, is pegged at the upper limit of 10, whereas $q_{\rm out}$ and $R_{\rm br}$ cannot be constrained. Thus we fix $q_{\rm out}$ at 3 as expected in the Newtonian case, and $R_{\rm br}$ at 10 $r_{\rm g}$. The model yields a broadly similar solution to the lamp-post model (see Table~\ref{tab:tab1}), also prefers a rapid black hole spin and a inner disk radius close to the ISCO. The reflection fraction in {\tt relxillCp} is defined differently from the {\tt relxilllpCp}, and the values are not directly comparable \citep[for a detailed explanation, see][]{dauser16}. The extremely high emissivity index makes the model highly sensitive to the position of the inner radius, resulting in much tighter statistical constraints on $a$ and $R_{\rm in}$. However, the extreme value of the emissivity index could be an indication that the broken power-law is an overly simplified description of the complicated emissivity profile. Also, although this results in a better fit, we note the main difference of Model 3 and Model 4 lies in the high energy tail (see Figure~\ref{fig:fig3}, left panel), where FPMA and FPMB data do not match perfectly, indicating the difference of the two models is close to the instrumental calibration accuracy. Therefore, we focus our discussion on the results from Model 3.

%%%%%%%%%%%%%%%%%%%%%%%%%%
\section{DISCUSSION AND CONCLUSIONS}
%%%%%%%%%%%%%%%%%%%%%%%%%%
We have performed a spectral analysis of the \nustar\ observation of the recently discovered black hole binary candidate \maxi\ in the bright hard state. Spectral fitting with relativistic reflection models measures a high black hole spin of $a>0.84$ and a small inner disk radius $R_{\rm in}<2.01~r_{\rm ISCO}$. It supports that the inner edge of the accretion disk remains close to the ISCO at the bright hard state of \maxi, the result is independent of whether a lamp-post coronal geometry is assumed or not. We measure a high reflection fraction $R_{\rm ref}=1.55$, which can be self-consistently described in the lamp-post model as the result of strong light bending effects near the black hole. Our spin constraint is consistent with the preliminary results reported by a \nicer\ observation in the intermediate state \citep{nicer17}. With basic properties of the binary system unknown (i.e., distance, black hole mass, orbital period), the exact Eddington ratio at the time of the observation is unclear. Assuming a distance of 8~kpc as the source direction is close to the Galactic center, we measure the source luminosity $L_{\rm 0.1-500~keV}\sim6\times10^{38}$~erg~s$^{-1}$, which is about half the Eddington luminosity for a 10~$M_{\sun}$ black hole \footnote{$L$ here is about the highest observed among black hole binaries in the hard state, and a few tens times higher than that in the hard state of Cygnus X-1.}.

In addition, we measure a lamp-post height of $h=7.2^{+0.8}_{-2.0}~r_{\rm g}$ and an electron temperature of $kT_{\rm e}=19.7\pm0.4$~keV, suggesting a compact and relatively cool corona. Low coronal temperatures, or equivalently low values for the high-energy cutoff, have so far been robustly measured in several black hole binaries during bright hard states with \nustar\ \citep[e.g.,][]{miller13, miller15}. Evidence has been found previously that the coronal temperature decreases during the rising phase of the hard state \citep[e.g.,][]{join08,motta09,gar15} and the high-energy cutoff disappears at the time of state transition \citep[e.g.,][]{bel06}, which signals a dramatic changes in the coronae. 

A weak narrow Fe K$\alpha$ component is required for an adequate fit of the \nustar\ data, which can be well described by a moderately ionized unblurred reflection model {\tt xillverCp}. A high resolution systematic study of Fe lines in X-ray binaries using \chandra/HETGS reveals that narrow Fe fluorescence emission is ubiquitous in high-mass X-ray binaries, but rare in low-mass X-ray binaries \citep{torr10}. It is unclear where the moderately ionized distant reprocessing material is located in \maxi. The best-fit absorption column density $N_{\rm H}=8.2^{+0.3}_{-0.6} \times 10^{22}$~cm$^{-2}$ is higher than the expected value for the Galactic absorption column, $N_{\rm H, Gal}=1.43\times10^{22}$~cm$^{-2}$ \citep{kalb05}. The extra obscuration is most likely intrinsic, which might be an indication of a complicated structure for the binary system. It is possible that the ionized reflection comes from the outer regions of a flared disk or the titled outer part of a warped disk. Ionized narrow Fe K$\alpha$ lines could also be produced in disk winds \citep[e.g.,][]{miller15wind, king15}, but is less likely in this case as there is no evidence for absorption lines and source is measured to be viewed at high inclination.

In light of the detection of QPOs reported in later observations \citep{nicer17,mere17}, we perform a search for possible QPOs in the frequency range of 0.1~Hz to 500~Hz in the \nustar\ power spectra, but obtain no significant detection, which is not uncommon considering the transient nature of the QPO behavior.

\acknowledgments{
We thank the referee for constructive comments that improved the paper. D.J.W. acknowledges support from STFC Ernest Rutherford Fellowship. J.A.G. acknowledges support from NASA Grant No.~80NSSC17K0515 and Alexander von Humboldt Foundation. This work was supported under NASA contract No.~NNG08FD60C and made use of data from the \nustar\ mission, a project led by the California Institute of Technology, managed by the Jet Propulsion Laboratory, and funded by the National Aeronautics and Space Administration. We thank the \nustar\ Operations, Software, and Calibration teams for support with the execution and analysis of these observations. This research has made use of the \nustar\ Data Analysis Software (NuSTARDAS), jointly developed by the ASI Science Data Center (ASDC, Italy) and the California Institute of Technology (USA).

\bibliographystyle{yahapj}
%\bibliography{maxi.bib}

\end{document}